\documentclass[12pt]{iopart}
\begin{document}
\title{A Derivation of Einstein Gravity without the Axiom of Choice: Topology Hidden in GR}
\author{M.\ Spaans}
\address{Kapteyn Institute, University of Groningen, 9700 AV Groningen, The Netherlands; spaans@astro.rug.nl}
\begin{abstract}
A derivation of the equations of motion of general relativity is presented that
does not invoke the Axiom of Choice, but requires the explicit construction of a choice function
${\bf q}$ for continuous three-space regions.
The motivation for this (seemingly academic) endeavour is to take the
background independence intrinsic to Einstein gravity one step further, and to assure that both
the equations of motion and the way in which those equations of motion are derived
are as self-consistent as possible. That is, solutions to the equations of motion of general
relativity endow a three-space region with a physical and distinguishing geometry in four-dimensional
space-time. However, in order to derive these equations of motion one should first be able to choose
a three-space region without having any prior knowledge of its physically appropriate geometry.
The expression of this choice process requires a three-dimensional topological manifold $Q$
to which all considered three-space regions belong, and that generates an equation of motion whose
solutions are ${\bf q}$. These solutions relate the effects of curvature to the source term through
the topology of $Q$ and constitute Einstein gravity. $Q$ is given by $2T^3\oplus 3(S^1\times S^2)$,
two three-tori plus three handles, and is embedded in four dimensions.
This points toward a hidden topological content for general relativity, best phrased as:
$Q$ and ${\bf q}$ provide a structure for how to choose a three-space region irrespective
of what geometric properties it has, while at the same time $Q$ and ${\bf q}$ determine that only GR
can endow a three-space region with those geometric properties.
So choosing a three-space region through an explicit choice function actually leads to Einstein gravity.
In this sense, avoiding the Axiom of Choice allows one to gain physical insight into GR, and perhaps into
the nature of physical laws. Possible links with holography are pointed out.
\end{abstract}

\section{Introduction}

General relativity (GR) is one of the most succesful and beautiful theories
ever constructed. It is a geometric theory, even though it predicts the
occurrence of topologically non-trivial structures
like black holes. Given the well motivated expectation of a multiply-connected
quantum foam on the Planck scale, it has been recognized that
the lack of any topological information in Einstein gravity is a hurdle in
its extension to quantum gravity. Conversely, an appealing property of Einstein gravity is its
independence of a fixed background, i.e.\ it is fully dynamical with
respect to the interactions between geometry and matter. The importance of background independence
is well recognized in the literature and is considered a valuable asset for any quantum theory
of gravity\cite{1}.
For instance, the loop quantum gravity approach of Ashtekar\cite{2,3} is explicitly background independent,
while string theory does not come with this attribute\cite{1}. The latter may change of course.

On the pure mathematical side, the Axiom of Choice (AoC) is well known in set theory
and expresses the intuitively obvious fact that one can always choose, for a collection $C$
of non-empty sets $S$ and a choice function $f$, a member from a set $S$, so that $f(S)\in S$. For finite sets
the AoC actually follows from the other axioms of set theory, but for infinite sets like the real
numbers (in multiple dimensions)
a definite choice of an element from each set $S$ does not exist unless the set is assumed to be well-ordered, which
is equivalent to adopting the AoC. When one invokes the AoC it is
then not necessary to actually construct a choice function since the assumption of its existence suffices.
Most mathematicians accept the axiom of choice
and those that do not belong to the camp of the constructivists. The fact that the AoC is
logically independent of the other axioms in Zermelo-Fraenkel set theory (ZF)
implies that one can never run into a logical contradiction when using the AoC\cite{4,5}.
Note that the combination of ZF and the Generalized Continuum Hypothesis (GCH) implies the AoC,
even though the GHC and the AoC are both independent of ZF. Since the GCH is a stronger constraint
than the AoC, attention is restricted here to the latter. There are many places where the
use of the AoC is required, i.e.\ when ZF alone is not sufficient in a proof but ZF and the AoC is.
Amongst them are set and measure theory, general topology,
algebra and functional analysis, most notably the theorem that every Hilbert space has an orthonormal basis.
The literature on the AoC is very rich\cite{4,5}, but for the purpose of this work it is sufficient to appreciate
that the AoC is not, but that a constructivist approach (i.e.\ explicitly determining $f$) is adopted.

This paper attempts to stay close to the standard derivation of GR, i.e.\
gravity is expressed through space-time curvature, and to remove any implicit dependence on the axiom of choice.
One might feel that this is a purely academic exercise. However, the crucial point
is the subtle difference between background independence, which GR
possesses, and self-consistency, which should apply both to the derived equations of
motion and to the way in which those equations of motion are derived.
That is, solutions to the equations of motion of general relativity endow a three-space region with a physical
and distinguishing geometry in four-dimensional space-time. This geometry can be used to define a choice function,
thus avoiding the need for the AoC. However, in order to derive those equations of motion one should first choose
a three-space region without having any prior knowledge of its appropriate geometry. This in turn would require the
AoC, which this paper attempts to repair.

The paper is organized as follows.
Section 2 presents the construction of the topological manifold $Q$.
Section 3 presents the resulting equations of motion that govern the
the choice function and links them to Einstein gravity.
Section 4 contains the discussion. The Einstein summation convention is used.

\section{The Construction of a Choice Process or Socks versus Shoes}

Since Einstein gravity is defined by local differential equations, even
the first pencil stroke leading to the GR equations of motion requires one to
consider a small (infinitesimal) three-space region of a four-dimensional manifold.
Because the region is infinitesimal, one can take it to be a simply connected region, i.e.\
one assumes that the region has no prior properties other than continuity. The {\it assumption}
of continuity, although plausible, is a necessary one.
Still, one is immediately confronted with the task of choosing such a three-space region.
At this point it is natural to think of a collection that contains all possible
topologically distinct manifolds as sets of simply-connected regions,
and the task to choose a simply-connected three-space region from each set.
For self-consistency of the procedure one can then best ask the following question:
Is it possible, for a connected set of three-space regions (in a four-dimensional space-time),
to choose a region from that set through a choice
process that is defined solely in terms of the three-space regions in the space-time itself?
If such a construction can be found then it is self-contained and does not require the AoC.

The starting point is formally a set $S$ that contains objects. These objects are
taken to be infinitesimal space-time regions, or one can even think of points for simplicity.
Note then that, in order to choose an object $x$, one needs to {\it select} it and to
{\it distinguish} it from another object $y$.
That is, one does not know, a priori, whether two selections
have not just singled out the same object. This is basically
the difference between having an infinite collection of socks versus one of shoes in Russell's apt
quote on the necessity of the AoC. Thus, to avoid the AoC one needs a distinguishing quality.
Clearly, shoes come with a left-right ordering, but it is only GR that can provide any ordering,
i.e.\ their intrinsic geometry, for the three-space regions in this work.
In this analogy, the choice function provides a means to assign geometric properties (turns socks into shoes),
but it is not the geometry itself (the shape of a shoe).
In any case, with this definition of choice one can proceed to define selection and distinction operations
through topological constructs, the motivation being that topology naturally allows one to
work with continuous, but still discrete, building blocks, which is intuitively appealing since
there is no fixed geometry provided by GR yet.

First, the intersection of two loops, $S^1\times S^1$, build from objects in $S$, facilitates selection of their
intersection $x$. For three-space regions the loops are solid rings (donuts), but for simplicity the one-dimensional
symbol $S^1$ is used since the three-space region is infinitesimal and should possess a point-wise (continuous) limit anyway.
Clearly, the dimensionality of the space one works in is crucial.
Three spatial dimensions are dictated by GR since it deals with the dynamics of three-space. In general three
spatial dimensions are very natural here since the two loops can then be non-trivially linked. In four dimensions
one can find a homeomorphism (a map that is continuous, is bijective and has a continuous inverse)
to unlink them, but in three dimensions
they form a single entity. Closed loops are desirable here to avoid boundaries, and the need to
specify (choose) conditions on them.

Second, a two-sphere, $S^2$,
with $x$ inside it then allows distinction, $y\ne x$, for $y$ outside of $S^2$ and inside its own two-sphere.
That is, two three-space regions are distinct if they can be seperated by two-sphere boundaries.

Third, the intersection region of two solid loops can be homotopically contracted to two crossing line segments that
locally span a plane because this does not change the overall topology, although it sacrifies the existence of a
continuous inverse. The also homotopically contracted selected object $x$ then lies on that crossing.
This homotopic allowance thus constitutes an implicit selection of two of the three spatial dimensions.
This argument may seem contrived, but the resulting three-space region is perfectly reasonable and not
excluded a priori on physical grounds, e.g.\ if effective dimensional reduction plays a role\cite{6}.
So for self-consistency of the choice function
the selection procedure needs to be implemented for all three possible pairs of spatial directions
(i.e.\ $x-y$, $x-z$ and $y-z$). This yields the three-torus, $T^3=S^1\times S^1\times S^1$.

Fourth, the solid loops and two-spheres used in the construction of the
choice function need to be selectable themselves as well to render the choice process self-contained globally.
Therefore, the distinct three-space regions $x$ and $y$, inside their individual $S^2$ spheres, need to be
connected by a handle, yielding $S^1\times S^2$ embedded in four dimensions, so that one can travel from one two-sphere
boundary to the other.
This in turn allows one to use a distinct three-space region itself in its {\it own} selection by looping it
through the handle back onto $T^3$, and thus one can incorporate any three-space region anywhere in a, thus linkable, solid loop.
This self-coupling in $3+1$ dimensions is the essential step in the construction of a self-consistent
choice structure, and all results below derive from it. Since at this point one deals with the homeomorphims
of a topology only, there is no clear sense of causality. Hence, {\it because of} the absence of an a priori geometry,
one can freely loop through the fourth dimension ('time') of the embedding space. This point will be re-addressed.

Finally, one just needs to consider {\it all} pairs of three-space regions $(x,y)$, i.e.\ any three-space
region needs to be distinguished with respect to all other regions.
This yields the connected, through three-ball surgery, sum of two (one for $x$ and one for $y$) three-tori
and three (for the $x-y$, $x-z$ and $y-z$ intersections) handles.
The resulting structure, denoted $Q$, is a three-dimensional topological
manifold $2T^3\oplus 3(S^1\times S^2)$ embedded in four dimensions, where $\oplus$
inicates the connected sum.
$Q$ can itself select and distinguish all three-space regions that it contains through all
the homeomorphisms that it allows, although there is not yet a way to label all the homeomorphisms.
The latter requires GR.

A priori, there is nothing physical about $Q$. It is merely a means to pick out all the different three-space
regions in a four-dimensional space.
Still, the logic that leads to $Q$ holds for any physical property of the three-space regions.
Given that there is no metric at this stage, one cannot assess the
sizes of different regions, i.e.\ one is dealing with socks because there are none of the physical properties yet
that define Einstein gravity.
Now, GR provides a measure of (a geometry for) different three-spaces and this changes the socks
into shoes, so to speak.
For instance, one can then order infinitesimal three-geometries using time or curvature.
Therefore, Einstein gravity by itself does not require the AoC and it is indeed only the derivation of its equations
of motion that should concern one. Therefore, the aim is now to derive equations of motion for a field
${\bf q}$ on $Q$ that lead to a clear metric rule (Einstein gravity), i.e.\ ${\bf q}$ identifies the geometries
and thus removes any dependence on the AoC.
The end result is then GR, but enriched with a (topological) structure that allows the derivation of its
equations of motion without invoking the AoC.

It is useful at this point to define operators $s_1$ and $s_2$ for the actions of selection and distinction,
respectively, rather than the underlying topological objects $S^1\times S^1$ and $S^2$.
That is, a choice is made by letting $m\equiv s_2s_1$ act on ${\bf q}$ from the left, $s_2s_1{\bf q}$.
For later use, the operator $n\equiv s_1s_2$ is defined as well.
One can then make the step to operations on a geometry by
the transformations $s_1\rightarrow {\bf X}\cdot$ and $s_2\rightarrow {{d}\over{d{\bf X}}}$ for
operators locally defined through a field ${\bf X}[x]$ on $Q$, e.g.\ the metric, that expresses the geometry.
One of course finds the commutator $[s_2,s_1]=1$. The interpretation of this commutator is
obvious. It matters whether one first selects an object and then distinguishes it by comparison with
some other object or that one selects this object after it has undergone some process that distinguishes it.
Obviously, $s_1$ and $s_2$ express, respectively, the need to assess the value of, and the change in ${\bf q}$ with, a
particular property used to describe a three-space region. There is no quantum mechanics here a priori, but merely the
observation that the choice process imposes a discretization of three-space.

As mentioned earlier, there is no clear sense of causality in the topological
manifold $Q$ and the homeomorphisms that it enjoys, and the same applies to (the time labelling of) the
commutation relation between $s_1$ and $s_2$. This is to be expected, since
$Q$ merely serves as a means to choose a three-space region. The solutions to the equations of motion for ${\bf q}$
derived below should provide a temporal order for the chosen three-space regions and a geometry, i.e.\ causality
enters through GR.
Thus, any three-space region $x$ is viewed as a field value ${\bf q}[{\bf X}(x)]$ and
$s_2s_1{\bf q}={\bf q}+{\bf X}{{d{\bf q}}\over{d{\bf X}}}$, i.e.\ one chooses ${\bf q}$ and also selects the
change in the chosen object, rendering the choice process self-consistent.
One may wonder why the operator $m$ is to be favored over $n$. The derivation of the equations of motion below will
show that it is not, and that only the product $nm=mn$ is pertinent.
Note the slight abuse in notation above by using the same symbol $x$ for the coordinates of an object and the object itself.
Well defined derivatives of ${\bf q}$
require more than just continuity and the equations of motion derived from $Q$ below will specify these necessities.

\section{The Equations of Motion or Identifying a Three-Space Region with GR}

\subsection{General Form or $Q$ Defines Possible Interactions}

To get to dynamical equations for ${\bf q}$ and to recover GR, note that
a three-torus defines four topologically distinct paths between any two points inside that
three-torus. This is easy to see because an individual three-torus contains three loops, which one can
use to detour over. Any path is locally $R^3$, but the three closed paths involving the $S^1$ loops are
not simply connected.
Now, the interactions present in the equations of motion for ${\bf q}$ must follow directly from
the topological structure of $Q$ if one does not wish to make additional choices.
So one should follow exactly the construction of $Q$, but now as experienced by the
field ${\bf q}$ defined on it. Under homeomorphisms, independent coordinate directions can be made to
coincide with the paths $\lambda$.
One is lead then to a single index, denoted by $\lambda =1..4$, equation because of the four
topologically distinct paths on $T^3$. The field $q_\lambda$ on a three-space region $x$
is part of some path $\lambda$ in $Q$ and is selected locally through
another crossing path, say $\mu$. The ${\bf q}$ field is defined on both these paths.
The two intersecting solid $S^1$ loops that define this selection can themselves be
distinguished as two single loops, denoted by case $B$, or
one double loop, denoted by case $F$, on the two three-tori.
The important point here is that one has two distinct ways in which to choose a three-space
region through ${\bf q}[{\bf X}]$:
$$s_2(s_2s_1){\bf q}={\bf X}{{d^2{\bf q}}\over{d{\bf X}^2}}+2{{d{\bf q}}\over{d{\bf X}}}.\eqno(1)$$
That is, either one selects and distinguishes two different regions and in turn selects
their comparison (case $B$, first term right hand side of (1)) or one distinguishes two selections made
through intersecting loops in their entirety (case $F$, second term right hand side of (1)).
Thus, one obtains a scalar ($S^1\times S^1$ is closed) second order derivative operator on
$[x\in S^1\subset S^2]\times [x\in S^1\subset S^2]$ for case $B$, called $D^2$.
Similarly, one finds a first order derivative operator on $[x\in S^1\times S^1]\subset S^2$
for case $F$, called $\delta$.
These operators are still unspecified, but they define ${\bf q}$'s dynamical interactions,
being $q_\lambda D^2 q_\mu$ and $q_\mu D^2 q_\lambda$ as well as $q_\lambda \delta q_\mu$ and $q_\mu \delta q_\lambda$.
It is obvious that under $B$ one distinguishes (counts)
all selections that the individual $S^1$ loops define, and there are arbitrarily
many of those, while under $F$ one
distinguishes only one object, the two intersecting $S^1$ loops themselves. Clearly, the
latter case allows a natural interpretation in terms of the Pauli
exclusion principle for fermions. Case $B$ then refers to bosons.
This is relevant for GR in the sense that both case $B$ and $F$ should lead to
equations of motion for Einstein gravity.

Furthermore, interactions that ${\bf q}$ experiences must in turn be selected, along any path and
by ${\bf q}$ itself, for self-consistency.
This is achieved by looping ${\bf q}$ through a handle, as argued above.
By now, the operator one is working with to derive the equations of motion is $s_1s_2s_2s_1=nm$.
Note then that this operator, and thus the equations of motion that derive from it, is independent of the
order in which one chooses because
$$nm{\bf q}={\bf X}^2{{d^2{\bf q}}\over{d{\bf X}^2}}+2{\bf X}{{d{\bf q}}\over{d{\bf X}}}=mn{\bf q},$$
so that $[m,n]=0$.
Finally, the {\it differences} between these ${\bf q}$ self-interactions, e.g.\ terms like $q_\lambda D^2 q_\mu$ - $q_\mu D^2 q_\lambda$,
when summed over all paths, must then be zero.
That is, one self-consistent choice order is as good as the other and thus are both equivalent.
This way, it does not matter whether one self-selects with $q^\lambda$ or $q^\mu$. So
$$q^\lambda [q_\lambda D^2 q_\mu - q_\mu D^2 q_\lambda ]=0=-0=q^\mu [q_\lambda D^2 q_\mu - q_\mu D^2 q_\lambda ]$$
and
$$q'^\lambda [q'_\lambda \delta q'_\mu - q'_\mu \delta q'_\lambda ]=0=-0=q'^\mu [q'_\lambda \delta q'_\mu - q'_\mu \delta q'_\lambda ],$$
where the equality to zero is imposed and one can relabel $\lambda\leftrightarrow\mu$.
For case $B$ and $F$, i.e.\ for the two distinct ways of chosing, ${\bf q}$ cannot be the same object,
which is indicated by a prime, ${\bf q'}$, for case $F$ above.
One finds then, for some path $\lambda$, that
$$q^\mu [q_\lambda D^2 q_\mu -q_\mu D^2 q_\lambda ]=0,\eqno(2B)$$
$$q'^\mu [q'_\lambda\delta q'_\mu -q'_\mu\delta q'_\lambda ]=0.\eqno(2F)$$
Indices are raised in (2) through, the still unspecified, metric tensor ${\bf g}$.
One has, equally unspecified, derivatives
$\nabla^{B,F}_\mu$. For case $B$, $D^2\equiv \nabla^{B\mu} \nabla^B_\mu$, and
for $F$, $\delta\equiv i\gamma^\mu \nabla^F_\mu$ with $\gamma^\mu$ the Dirac
matrices (indices suppressed, i.e.\ $\Sigma_\alpha \gamma^\mu_{\lambda\alpha}\nabla^F_\mu q_\alpha$ implied).
The definition of $\delta$ follows from the fermionic interpretation of case $F$.
In this then, ${\bf q'}$ is a Grassmann, anti-commuting, variable, with ${\bf q'}^2=0$.
This immediately eliminates the second term in equation $(2F)$.
Where $Q$ only needs continuity (or rather the existence of homeomorphisms), ${\bf q}$ requires well defined
second order derivatives. The latter requirement is not an external constraint because it is the topology of $Q$ that
leads to the equations of motion for ${\bf q}$.

For a simply connected region one should have locally, i.e.\ in flat space, that
$g^{\mu\nu}=\eta^{\mu\nu}$, for the Minkowski metric ${\eta^{\mu\nu}}$,
and $\nabla_\nu =\partial_\nu$, provided that the manifold $Q$ supports this metric
signature. The latter is immediate, since it is a statement of special relativity and $Q$,
by construction, is background-free. So information propagating
on $Q$ must do so in the same way, i.e.\ at the same speed, to render all observers identical.
The equations of motion $(2B)$ show this formally as well. Consider the general solutions
$D^2q_\lambda =b^\mu b_\mu q_\lambda$ with ${\bf q}\ne 0$ and ${\bf b}[\{{\bf q,q'}\}]\ne 0$. These solutions are
equally viable for $b^\mu b_\mu =0$, irrespective of the form of $D^2$. But then it follows that the
topological manifold $Q$ must support a seminorm for an in-product since $b_\mu\ne 0$. This allowance
of Lorentz symmetry is simply a gauge invariance of solutions to the equations of motion and results because
${\bf b}$ may depend on any of the solutions $\{{\bf q,q'}\}$ to the equations of motion (2), while remaining
a closed system. For example,
the solution $q_\lambda =b_\lambda =(t, x, y, z)$, yielding $D^2q_\lambda =b^\mu b_\mu q_\lambda =0$, works fine
for any metric signature $\eta$ in $D^2$ on flat space,
but a zero in-product leads to $\eta =-+++$ (all three spatial directions must have the same sign).
Consequently, $D^2$ and $\delta$ are forced to reflect a Lorentzian manifold.

Finally, further contraction with $q^\lambda$ and $q'^\lambda$
renders the left hand sides of (2) {\it identically} zero.
Any other result would have lead to an immediate failure of the self-consistency of the choice function.
Clearly, after following all paths through $Q$ all choice orders have been encountered, irrespective of
a law like gravity.
Also note then that the {\it form} of equations (2) is determined solely by the topology
of $Q$ and therefore is invariant under homeomorphisms (and thus diffeomorphisms). This implies the general
covariant description of any physical theory derived from (2). Of course, this is merely a statement of
mathematics, not physics.
One can conclude then that where general covariance provides a framework to study geometry independent of
coordinates, $Q$ and ${\bf q}$ provide a structure for how to choose a three-space region irrespective
of what properties GR can endow it with.

\subsection{Einstein Gravity or Degrees of Freedom from Topology}

After the above preliminaries, it is time to arrive at the equations of motion for GR, guided by the
topology of $Q$. One sees immediately, for $\nabla^{B,F} =\partial$, that
$D^2 q_\lambda=q_\lambda$ solves $(2B)$, while the (dimensionless) Dirac equation,
$\delta q'_\lambda =q'_\lambda$, solves (2F). For case $B$ it is natural then to
introduce a general source field ${\bf s}[\{{\bf q,q'}\}]$ for $D^2 {\bf q}={\bf s}$.
One merely breaks up the information provided by ${\bf q}$ into a dynamical part and a source part to bring
out the role of topology. The solution $D^2 q_\lambda =s_\lambda$ yields the constraint on the source
$$q^\mu q_\lambda s_\mu -q^\mu q_\mu s_\lambda =0,\eqno(3B)$$
with the obvious solution ${\bf q}={\bf s}$.
The interchange symmetry of the two identical three-tori in $Q$,
$T^3\leftrightarrow T^3$, then yields a new source object ${\bf T}[\{{\bf q,q'}\}]={\bf T'}[\{{\bf q,q'}\}]$,
for case $B$ {\it and} $F$, that involves two indices. Similarly, ${\bf q}$ transforms to a two-index object ${\bf G}$.
This is because 1) the loop topologies that express bosons
and fermions are both subsets of an individual three-torus and 2) the interchange symmetry forces one to consider
{\it pairs} of paths. These paths are denoted for simplicity by ($c_1$,$c_2$) and $S^1=c_1 c_2^{-1}$ must hold.
That is, as one travels from one three-torus to the other on $Q$, the interchange symmetry
moves one right back to the starting point through identification. Physically, one can exchange two separate {\it simply-connected}
three-space regions with source fields on them, each region being part of an individual three-torus, without changing
their mutual gravitational interaction. Therefore, one has a spin 2 field, with four dynamic degrees of freedom from
the four distinct paths on a three-torus in $Q$.
In this, the mathematical need for a parallel transport rule is obvious since the geometric choice rule that ${\bf q}$
constitutes after the source split-off may not change intrinsically from one three-space region to the next.
Furthermore, torsion occurs if two tangent vectors in a point, when regarded as small
displacements, do not close under parallel transport. Under homeomorphisms $S^1$ closure, irrespective of the
connection, can always be achieved, while the curvature of $S^2$ surfaces cannot be transformed away under
homeomorphisms.
So the Levi-Civita connection applies, given that $\nabla^B_\lambda g_{\mu\nu}=0$ (metric compatibility)
must hold. That is, the norm of a vector cannot change under parallel transport either since equation $(2B)$ has
solutions $D^2 q_\lambda =q^\mu q_\mu q_\lambda$ (${\bf b}$=${\bf q}$,
and similarly for different Grassmann fields under $(2F)$), and thus neither
${\bf q}$ nor $q^\mu q_\mu$ may change under parallel transport.

Still for case $B$, equation $(3B)$ is transformed,
for ${\bf s}\rightarrow {\bf T}$ and ${\bf q}\rightarrow {\bf G}$, to
$$G^{\alpha\beta}G_{\lambda\mu}T_{\alpha\beta}-G^{\alpha\beta}G_{\alpha\beta}T_{\lambda\mu} =0.\eqno(4B)$$
Solving $(4B)$ yields $G_{\mu\nu}=T_{\mu\nu}$ trivially because of the source split-off (${\bf s}$=${\bf q}$).
Of course, the properties of ${\bf G}$ must be found to actually reproduce GR. To this effect,
note first that ${\bf q}$ implicitly depends on ${\bf G}$ and that
${\bf q}[{\bf G}]={\bf q}[{\bf T}]={\bf q}[{\bf G}({\bf g})]$
is the quantity that contains all the distinguishing information about the three-space geometry.
So one should
always transform back to a ${\bf q}$ (or ${\bf q'}$ under $F$) form of the equations of motion
to confirm the validity of the found solution.
For case $B$ this is obvious. In general,
${\bf g}[{\bf q}]\rightarrow {\bf R}[{\bf g}({\bf q})]$, along $S^1$, as one combines all four paths on
a three-torus with one another,
i.e.\ one finds an object $R_{\mu\nu\kappa\lambda}$ with four indices, purely from the topology of $Q$.
This ${\bf R}$ must yield an object $G_{\mu\nu}[{\bf R,g}]$ that is
restricted to $S^1$ and that has only two indices, one for $c_1$ and one for $c_2$, because of $Q$'s exchange
symmetry. On $S^1$ any order along $c_1$ and $c_2$ can be taken, so one has that $T_{\mu\nu}=T_{\nu\mu}$ and
that $G_{\mu\nu}=G_{\nu\mu}$.

For case $B$, ${\bf G}$ can now be found given that ${\bf R}$
must be fixed by the {\it difference} $q^\mu_{c_2}-q^\mu_{c_1}$, i.e.\ the distinction
between the two paths finally allows one to choose independent of the AoC. So, one has that
$$q^\mu_{c_2}-q^\mu_{c_1}=q_0^\kappa R^\mu_{\kappa\lambda\nu}c_1^\lambda c_2^\nu,\eqno(5B)$$
for the displacements $c_1^\lambda ,c_2^\nu$ along $S^1$ and an arbitrary
point $0$ (with ${\bf s}$=${\bf q}$). Indeed, ${\bf R}$ is precisely the Riemannian curvature.
This is the relationship between the choice function and the metric in GR:
The four-component tensor field ${\bf q}$
relates the effects of curvature to the source term through the topology of $Q$.
The maximum order of derivatives of ${\bf g}$ in ${\bf G}[{\bf R,g}]$ is two because there are two
distinct paths that enter ($s_2s_2$ in equation (1)), and ${\bf G}$ is linear in these second derivatives because
the interchange symmetry acts through $S^1$ (and not some higher power thereof).
The constraint $\nabla^{B\mu} G_{\mu\nu}=0$ must hold under
$T^3\leftrightarrow T^3$ since $\nabla^B{\bf G}[c_1]=\nabla^B{\bf G}[c_2]=\nabla^B{\bf G}[c_1^{-1}]=-\nabla^B{\bf G}[c_1]=0$,
i.e.\ under homeomorphisms one can always deform, and effectively interchange, the individual curves making up $S^1$
such that the net change {\it along a path} in the object ${\bf G}$, generated by $T^3$ exchange, is zero.
This is crucial since $\nabla^B{\bf G}=0$ fixes the dependence of ${\bf G}$ on the Ricci tensor and curvature scalar in
the right way, yielding the Einstein tensor, and gives $\nabla^B{\bf T}=0$.

Also, for case $F$, under $\delta {\bf q'}={\bf s'}\rightarrow {\bf T}$ and ${\bf q'}\rightarrow {\bf G'}$,
the left hand side of equation $(2F)$ transforms analogously to case $B$ and becomes
$$G'^{\alpha\beta}G'_{\lambda\mu}T_{\alpha\beta}=0.\eqno(4F)$$
One has identical results as above since ${\bf G'}={\bf T}$ solves $(4F)$,
after transforming back to ${\bf q'}$ form, through ${\bf q'}[{\bf G'}]{\bf q'}[{\bf T}]=0$.
Thus ${\bf G'}={\bf G}$, under zero torsion, and fermions and bosons see the same geometry,
${\bf q'}[{\bf G'}]={\bf q'}[{\bf G}]$ for case $F$ and
${\bf q}[{\bf G'}]={\bf q}[{\bf G}]$ for case $B$.

Most importantly, the weak equivalence principle (i.e.\ the one closest to Einstein's original formulation)
follows naturally from the solutions to (4) since the only sense of mass, as a distinct property, that $Q$ possesses
is through the curvature of the handles. That is, one can associate some measure of mass with a region enclosed by an
$S^2$ surface, but it is always related to curvature.
The same notion of mass must then hold for ${\bf T}$ through the source split-off, i.e.\ ${\bf q}={\bf q}[{\bf G}]$ (and
similarly for ${\bf q'}$) while ${\bf G}={\bf T}={\bf T}[\{{\bf q,q'}\}]$, closely following Mach's principle.
So mass is an expression of topology here and the gravitational/geometric motion
of a small body is independent of its composition.
In all, this constitutes the (dimensionless) content of Einstein gravity.
The usual procedure of taking the Newtonian limit needs to be followed to recover the constant $8\pi G$ in front
of ${\bf T}$, since $Q$ possesses no intrinsic scale.
Of course, the Cauchy problem needs to be solved to get to a metric, but here the topology of $Q$ also
plays a role. It provides a compact three-space, e.g.\ the (isotropic) three-torus or a handle, as a hypersurface.
Finally, torsion can be present if one considers the closed loops inside a $T^3$ and the spin $1$ and spin $1/2$
(under $B$ and $F$) states that they represent. Although Einstein-Cartan theory, e.g.\ spin-orbit coupling, is beyond
the scope of this derivation, it is obvious that $Q$ does possess a topological analog because the three-tori are
not simply connected.

\section{Discussion}

The basic result of this work is that choosing a three-space region through an explicit choice function
leads to Einstein gravity.
There further appears to be a connection between the topology of $Q$ and holography.
The $S^2$ surfaces required by the distinction operator $s_2$ in the choice function clearly
define what is internal and external about observational reality, i.e.\ what is accessible information.
As shown by 't Hooft\cite{6} and Susskind\cite{7} this goes a long way toward understanding what properties
black holes, and in general event horizons, have. Conversely, it is ${\bf q}$ that allows one to assign specific
information to a three-space region. One is thus lead to conclude that by its very topology $Q$ limits
${\bf q}$'s information content to the closed boundaries of nested three-space regions only. A black hole is then
the breakdown of homotopic contractibility, creating an innermost closed surface.

One might feel that the presence of one fixed topology constitutes
a fixed background. This is not the case since the topology of $Q$ is
the result of a self-consistent implementation of a choice process of continuous
three-space regions. Of course, for objects more basic than continuous spaces a further generalization
of the topological manifold $Q$ would be necessary. The presence of the handles strongly resembles the picture of Wheeler's
space-time foam. In any case, there appears to be a rich topology hiding in Einstein gravity that can be
brought out by explicitly constructing a choice function rather then invoking the AoC. One might speculate that
the logic followed for GR applies to any derivation of equations of motion. If so, then $Q$ should generate more
physical solutions to (2) through the symmetries that its topology possesses. Avoiding the axiom of choice, i.e.\
constructivism, may then help one to derive physical laws in general.

\medskip\noindent
{\it Acknowledgments}\ Discussions with Lee Smolin are greatly appreciated.

\end{document}